# Point Cloud Rendering after Coding: Impacts on Subjective and Objective Quality

A. Javaheri, C. Brites, *Member,* IEEE, F. Pereira, *Fellow,* IEEE, J. Ascenso, *Senior Member,* IEEE

*Abstract*— Recently, point clouds have shown to be a promising way to represent 3D visual data for a wide range of immersive applications, from augmented reality to autonomous cars. Emerging imaging sensors have made easier to perform richer and denser point cloud acquisition, notably with millions of points, thus raising the need for efficient point cloud coding solutions. In such scenario, it is important to evaluate the impact and performance of several processing steps in a point cloud communication system, notably the degradations associated to point cloud coding solutions. Moreover, since point clouds are not directly visualized but rather processed with a rendering algorithm before shown on any display, the perceived quality of point cloud data highly depends on the rendering solution. In this context, the main objective of this paper is to study the impact of several coding and rendering solutions on the perceived user quality and in the performance of available objective assessment metrics. Another contribution regards the assessment of recent MPEG point cloud coding solutions for several popular rendering methods which was never presented before. The conclusions regard the visibility of three types of coding artifacts for the three considered rendering approaches as well as the strengths and weakness of objective metrics when point clouds are rendered after coding.

## I. Introduction

Nowadays, emerging 3D visual representations allowing more immersive experiences compared to the classical 2D images or videos are attracting much interest. In fact, a new wave of multimedia applications are now possible, from geographical information systems, and virtual and augmented reality to cultural heritage and free viewpoint broadcasting, motivated by the recent advances in 3D acquisition systems [1]. In this context, point clouds are becoming an important 3D visual representation format of the real world due to the availability of several acquisition devices (from range sensors to multi-camera arrays) as well as efficient coding solutions and rendering techniques. A point cloud (PC) is a set of 3D points represented by their 3D coordinates and associated attributes, such as color, normals and reflectance. PCs can be classified with respect to their temporal evolution. While static PCs correspond to a single time instant, dynamic PCs correspond to a PC evolving along time, thus corresponding to a sequence of static PC frames. Also, progressive PCs correspond to large-scale PCs that are not consumed all at once and thus are made from complementary parts of a visual scene; these parts are static PCs that differ both spatially and/or temporally (often used in autonomous driving).

To represent the visual scene with high fidelity, a PC can have several millions or even billions of points, which results in a large amount of data that needs to be efficiently stored and transmitted. Thus, coding technologies are essential to deal with the huge amount of data that PC acquisition devices can generate. The coding solutions already available [2]-[5] can be lossy or lossless and aim to reduce the PC representation bitrate while keeping the data fidelity as high as possible. Following the demands by the industry, both the Joint Photographic Experts Group (JPEG) and the Moving Picture Experts Group (MPEG) standardization bodies recognized that the PC format can address future immersive multimedia applications and have initiated projects in the area of PC coding [6][7][8]. In January 2017, MPEG has issued a Call for Proposals on Point Cloud Compression (PCC) [9], targeting the efficient representation of static objects and scenes, as well as dynamic objects and real-time environments. After this call, two PC coding solutions have been developed, notably the so-called *Geometry-based Point Cloud Compression* (G-PCC) standard [10], for static and progressive acquired content and *Video-based Point Cloud Compression* (V-PCC) [11] standard, for dynamic content.

Naturally, PC quality assessment is fundamental to evaluate the performance of the several processing steps in PC-based applications, notably denoising, coding and rendering. Moreover, subjective quality assessment procedures and objective assessment metrics that can accurately evaluate the perceived quality, notably when PC data is compressed, are much needed. Both are critical to improve the final Quality of Experience (QoE) offered to the end-users, not only to monitor the quality of the experiences but also to allow the design and optimization of novel PC coding techniques.

PCs can be visualized on a variety of devices, such as 2D displays, head-mounted displays (HMDs), augmented reality devices and even on stereoscopic or multi-stereoscopic displays. However, independently of the type of display, PCs cannot be directly visualized and require a rendering technique to create the data that may be visualized; this can be seen as a post-processing step after decoding. Nowadays, there are multiple PC rendering approaches [12] [13] that may significantly influence the perceived PC quality in different ways. While there are several subjective and objective quality evaluation studies available in the literature, they do not use the same type of coding and rendering solutions as well as test conditions and thus, rather often, reach different conclusions. Therefore, it is critical to study the impact of different rendering approaches on the subjective and objective decoded PC quality.

On the other hand, many relevant past works on subjective and objective quality assessment [14]-[20] rely on simple coding solutions such as *octree pruning*, which are inefficient and produce a rather distinctive type of artifacts. However, more sophisticated and also more efficient lossy PC coding solutions

The authors are with the Instituto Superior Técnico and Instituto de Telecomunicações, 1049-001 Lisboa Portugal {email: alireza.javaheri@lx.it.pt, catarina.brites@lx.it.pt, fp@lx.it.pt, joao.ascenso@lx.it.pt } are now available, which produce decoded PCs with very different characteristics and artifacts. As an example, some PC codecs significantly increase the number of decoded points to hide coding artifacts, thus achieving a better perceived quality. This makes the subjective and objective quality assessment of PCs more complex, especially when more efficient coding and rendering solutions are considered.

While PCs have commonly two major components, geometry and color (or texture), this paper focus on the quality impacts of degradations on the geometry component of the point cloud representation. Geometric artifacts are very important for the final perceived quality since this type of degradations may reduce the realism of the decoded geometry, e.g. due to the appearance of holes and deformed and noisy surfaces, consequently leading to poorer user immersion. Fig. 1 shows an example of geometric artifacts, in this case associated to the MPEG G-PCC codec (original texture was used for recoloring), which clearly results in a rather low perceived quality. Despite its importance to the perceived quality, geometric degradations have not been addressed much in the literature before. Besides, while the geometry is an intrinsic component of the PC representation, the color attributes (which are optional) may not be available due to limitations in the acquisition process, e.g. PCs acquired by LIDAR only devices.

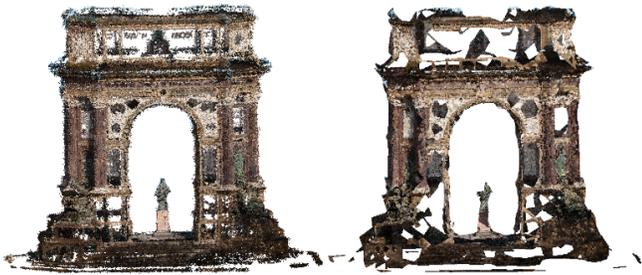

Fig. 1. *Arco Valentino* PC: left) original PC; and right) MPEG G-PCC decoded PC. Original texture used for recoloring.

In this context, the objective of this paper is to study in a subjective way, the impact of the different artifacts produced by state-of-the-art PC codecs for different types of rendering and assess objective PC geometry metrics in several scenarios. This is the first time that the rendering $\mathcal{R}$ and coding $\mathcal{C}$ processes, which play a major role on the final perceived quality, are jointly evaluated in this case for static point clouds. To be able to isolate and, thus, directly assess the impact of the geometric artifacts on the perceived quality, no color attributes coding was considered in this work. In this context, the main contributions are:

- **PC rendering after coding – subjective quality assessment:** Study of the subjective quality impact of multiple $(\mathcal{R}, \mathcal{C})$ combinations for relevant, lossy PC coding and rendering solutions. Moreover, the visibility of the distortions associated to each codec under different rendering scenarios will be analyzed. This first contribution is critical for the design of a suitable PC subjective assessment methodology, where a rendering solution must be chosen.
- **PC rendering after coding – objective metrics assessment:** Evaluation of the performance of available PC objective metrics for multiple $(\mathcal{R}, \mathcal{C})$ combinations, i.e. for different types of rendering and coding artifacts. This should allow understanding the strengths and weaknesses of available objective metrics as well as their scope of validity, i.e. for which conditions these metrics represent well enough the human perceived quality. This second contribution is critical for the design of more reliable PC objective metrics, notably for the evaluation of new PC coding solutions as well as associated techniques.
- **Rendered Point Cloud Quality Assessment Dataset**: Provision of the first public dataset of mean opinion scores (MOS) and corresponding PCs coded with relevant, lossy PC coding solutions. These PC codecs produce a distinctive set of artifacts that were not considered when the popular PC metrics were designed. This third contribution is particularly important for the young PC quality assessment community, since not many subjective studies are available, and from the ones available, none allows to assess the impact of the rendering process that is always performed after decoding.

This paper is organized as follows. Section II reviews the related work while Section III describes three key PC coding and rendering solutions, which are used for the following experiments. Section IV describes the subjective evaluation study along with some key conclusions. Section V aims to evaluate the most relevant PC objective metrics. Section VI presents some final remarks and, finally, Section VII presents challenges and proposes possible ways forward towards the advancement of this technical area.

## II. RELATED WORK

In the literature, there are several subjective and objective PC quality assessment methods and studies available. In [21], Zhang *et al.* designed a subjective test for colored PCs under different levels of degradation of both geometry and color. The quality degradations have been introduced by down-sampling the geometry and independently adding (synthetic) uniform noise for both color and geometry. The main conclusion was that human perception is more tolerant to color noise compared to geometry noise in PCs.

In [22], Mekuria *et al.* conducted the subjective evaluation of a PC codec based on geometry octree pruning and JPEG based attributes coding. The subjective evaluation was performed in a mixed reality system, combining coded PC data (acquired) and computer graphics generated 3D content. In the subjective test, the users could interact with the content by navigating a visual scene with an avatar. The system performance was globally assessed with a questionnaire addressing eight different quality aspects, notably realism, immersiveness and color quality. Two objective metrics (mean squared error based) were introduced to assess both the geometry and color qualities. However, the correlation between objective metrics and subjective results was not assessed.

In [23], Javaheri *et al.* performed the subjective and objective quality assessment of denoising algorithms for PC geometry. To introduce geometry errors in clean, reference PCs, impulse noise and Gaussian noise with three different strengths were added to represent three different perceptual qualities. Several outlier removal and regularization algorithms were applied to the degraded PCs. Noisy and denoised PCs were rendered on standard 2D displays by first applying a surface reconstruction method (i.e. PC converted to polygonal meshes). Also, several objective metrics were selected to assess their performance in a denoising context. In a later study [24], Javaheri *et al.* performed the subjective and objective quality assessment of the geometry of compressed PCs. In this case, the popular Point Cloud Library (PCL) octree and graph-based codecs, with two very different types of associated distortions, were used. The rendering was performed with a point-based representation, recoloring each decoded point with the original color attributes. In both these works, PCs were visualized on a 2D display and a *Double-Stimulus Impairment Scale* (DSIS) method was used for subjective quality assessment.

In [14], Alexiou *et al.* performed a subjective quality assessment study of the PC geometry for two types of degradation, octree pruning and Gaussian noise, thus generating PCs with different quality levels and artifacts. An augmented reality (AR) headset was used to visualize simple PC objects without color from different perspectives (user could move around the object). It was concluded that objective metrics could perform well for Gaussian noise but underperform for PCL-like compression artifacts. In [15] and [16], Alexiou *et al.* also performed a subjective test with the same data as in [14] and the same distortion types but visualized on a 30-inch 2D display. In both cases, color was not used and user interaction was allowed; a simple rendering method with unit size points was selected. In [15], the impact of adopting two different subjective test methodologies, *Absolute Category Rating* (ACR) and DSIS, was studied through their comparison. The DSIS methodology was found more consistent and with lower confidence intervals and, thus, it was used later in [16]. In [17], Alexiou *et al.* performed a subjective study to evaluate the PC geometry quality using an octree pruning based codec. Before rendering, a Poisson surface reconstruction algorithm was used to obtain a mesh from the decoded PCs. In this case, no interaction was allowed with the content and the subjective experiment also followed the DSIS methodology. It was found that most PC objective metrics have a low correlation with the subjective scores and the 3D surface reconstruction algorithm plays a crucial role on the subjective scores obtained.

In [18], Alexiou *et al.* performed two subjective tests to study the impact of visualization on the subjective quality assessment of PCs. The first test used a 30-inch 2D display and the second an AR headset. As before, geometric artifacts associated to octree coding and Gaussian noise were used. The test methodology was DSIS and interaction by users was not allowed. In any case, the correlation between scores obtained with different visualization devices was rather high, notably statistically equivalent for Gaussian noise.

In [19], Alexiou *et al.* conducted a subjective evaluation to assess the quality of compressed PCs rendered as mesh objects in several types of 3D displays, from passive stereoscopic to auto-stereoscopic displays. Geometry degradations in the form of octree pruning were evaluated in the absence of color. The results obtained with 3D displays have a strong correlation with the results obtained with 2D displays for the same content.

However, it was also found that the rendering method may play a significant role in this evaluation. Also, Alexiou *et al.* have benchmarked objective metrics for PC data represented by octree pruning and corrupted with Gaussian noise [20]. Both DSIS and ACR methodologies were used in separate sessions. It was found that the correlation between subjective and objective scores was low for distance-based objective metrics for octree-based compression artifacts, but better correlation performance could be achieved with metrics considering the normal at each point.

In [25], Christaki *et al.*, performed a subjective study for simple PCs, that were converted to meshes and coded with suitable open-source mesh codecs. While some of the test PCs are common with the PCs often used in previous subjective quality evaluation studies (e.g. *Bunny*) others were obtained with a platform designed for 3D human capture (with multiple Kinect devices). In [25], a variant on the pairwise subjective test methodology was used for evaluation with three stimuli presented simultaneously. Overall, three mesh codecs were considered, and content was displayed with a virtual reality (VR) application in a head-mounted display. They concluded that usual 3D mesh metrics have a low correlation performance in this scenario and the 3D mesh surface reconstruction method plays an important role. Finally, Dumic *et al.* presented in [26] the state-of-the-art on PC subjective quality evaluation as well as a summary on the available PC objective metrics.

In many of studies reviewed above, it is concluded that the rendering process, applied after decoding, can have a significant impact on the perceived quality by the users; however, there is no solid assessment or quantification of the differences between rendering methods. Moreover, realistic distortions produced by relevant coding solutions are not often used, e.g. compression artifacts have been artificially simulated by noise addition or coding solutions that are much less efficient, compared to the MPEG PC codecs. Also, all the previous studies on PC quality assessment do not follow a common set of test conditions, such as those defined by the MPEG and JPEG standardization bodies. Finally, many previous works just focus on a single type of objective metrics. These limitations and simplifications are overcome by this paper which precisely targets to study the impact of the rendering process on the perceived quality and objective metrics accuracy for recent, efficient coding solutions, under meaningful test conditions, for a wide range of objective metrics. This should guide future developments in the areas of subjective and objective PC quality assessment.

III. POINT CLOUD CODING AND RENDERING SOLUTIONS

This section first describes three popular rendering solutions, used later for subjective and objective assessment. Then, three well-known state-of-the-art PC codecs are reviewed, and the associated PC coding artifacts are characterized.

*A. Selected Point Cloud Rendering Solutions*

PC rendering is the process of producing a visual representation that can be consumed by users using an available display, e.g. conventional 2D, stereo, auto-stereoscopic, head-mounted displays, etc. [27]. Since it effectively selects the information to be seen, the rendering process has a significant impact on the quality perceived by the user. In this section, the rendering solutions selected for the experiments reported in this paper are briefly described.

Regarding PC rendering, there are two main approaches; the first, directly uses the PC data (point-based) while the second converts the PC data into another representation format, very commonly a surface, e.g. a polygonal mesh. The decision on the rendering approach to adopt mostly depends on the application requirements which may be very different.

The PC conversion to another representation format more rendering friendly may bring some information loss and, in some cases, it may not even be possible due to the complexity of the visual scene in terms of geometry or the low PC density. By directly rendering PCs, massive amounts of points can be visualized. Rather often, these PCs do not fit into the available memory and require special algorithms to stream, process and render only a small subset of the entire PC data. This is easier to perform with a point-based model due to the lower complexity associated to the rendering process in comparison to a polygonal mesh representation where surface reconstruction and interpolation are usually needed.

Independently of the rendering approach, a geometry shader with some primitives is employed to construct the final image shown to the user. In this context, the geometry shader is responsible for the creation of appropriate levels of light, darkness, and color within an image [28]. For PCs only with geometry, the shading is commonly performed with a single color; otherwise, color attributes are used for each point or vertex. Moreover, primitives are the simplest (atomic) elements that are combined to create the 3D impression of surface in the final displayed content.

*1) Point-based Rendering without Color Attributes*

Point-based rendering algorithms use a set of discrete points that may be irregularly distributed, simple rendering primitives and 3D/image space interpolation procedures to obtain a 2D image. The main advantage of this rendering approach is that it can achieve high levels of realism and is adequate for complex objects, such as trees, feathers, smoke, water, etc. In addition, point based rendering simplifies the rendering process and typically requires less memory and computational power due to the lack of connectivity information.

In this approach, simple and fast to render primitives are selected, such as circles, squares, spheres, cubes. Based on the PC density and distance to the virtual camera (zoom level), the size of the primitives can be manually or automatically adjusted to create the impression of a surface; in the automatic case, connectivity information between points is usually computed to determine the primitive size [24]. The definition of an appropriate size for the primitive is rather important to reduce the appearance of empty spaces (holes) between points (size too small) or aliasing artifacts (size too large). In this work, the primitive selected for rendering was a square because they are similar to the smallest element of a 2D image (pixels) and the point size was set to the minimum value able to fill the 3D space between points completely, thus avoiding holes.

Regarding shading, color attributes were not used, in order the impact of geometry distortions may be assessed without any additional component. The human visual system can easily and accurately derive the three-dimensional orientation of surfaces by using variations in the image intensity alone [29]. To obtain the normals, a (best fitting) plane was used as the local surface model and an automatic estimation for the neighborhood radius was used, as suggested in [17]. This automatic estimation helps to find a suitable radius as a too small radius may result in some points having an invalid normal and a too large radius may result into smoothed edges. By fitting a local surface, only the direction of the normal can be computed and, thus, the orientation of the normal was determined with the minimum spanning tree algorithm [30]. This type of rendering approach will be designated as *RPoint* in the following.

*2) Point-based Rendering with Color Attributes*

The second rendering solution is still point-based but uses also the available color attributes and thus for this reason, it will be designated as *RColor* in the following. In *RColor*, the *RPoint* rendering method is again applied but the point color attributes are used. This means that the surface is still represented with points and displayed with the same primitives but with the color obtained during the PC acquisition process. While the color attributes correspond to the real color of the objects, they are still influenced by the specific light conditions that have occurred during their acquisition. However, in the final rendered image, some colors can be interpolated, e.g. between points, to avoid aliasing. Moreover, since the captured color also conveys the object depth, it may mask some geometric distortions of the surface. On the contrary, distortions may be more visible at object boundaries, which give the user, the shape perception of the objects in the visual scene.

In this work, to isolate the impact of geometric distortions, the color attributes are not compressed and, thus, the original color is used to recolor the decoded PC. The recoloring process occurs when the number of points in the decoded point cloud is different (or the same) from the original number of points. The recoloring procedure uses the original color and performs a mapping of the original colors in the original positions to the decoded points positions. In this case, the vertex attribute transfer method available in MeshLab was used for the recoloring process. Moreover, in the adopted *RColor* rendering method, no relighting is performed to preserve as much as possible the color fidelity of the PC representation.

*3) Mesh-based Rendering*

The first step in the mesh-based rendering approach, hereafter designated as *RMesh*, is to create polygonal meshes with a surface reconstruction algorithm, such as the *Poisson Surface Reconstruction* [31]. This means that rendering is performed with a set of vertices along with their connectivity to obtain a closed surface very precisely defined.

The advantage of this rendering method is that, independently of the distance to the object (or scene) or the PC density, a seamless surface is obtained; this may not occur with point-based rendering since the quality is associated to the number of points describing the surface and the distance between the viewer and the object. The disadvantage of this rendering method is that it requires surface reconstruction, which usually removes high frequency geometric details [32] (smooth surfaces are obtained). It is important to note that surface reconstruction from complex surfaces is not always straightforward, it may not always be successful and can even require some user intervention. After surface reconstruction, the polygonal mesh needs to be rendered, usually with some shading algorithm [33][34]. There are several mesh rendering techniques performing shading, reflection, refraction and indirect illumination, and able to improve (when properly applied) the quality of the rendered data.

In this work, the procedure to reconstruct the surface proposed in [17] was followed. The *Poisson Surface Reconstruction* algorithm, available in the popular *CloudCompare* [35] software, was selected with default parameters. The estimation of the normal vectors was performed as for the *RPoint* solution; no color attributes were used to be able to directly assess the subjective impact of the geometric artifacts.

### B. Selected Point Cloud Coding Solutions

This section reviews some relevant and representative PC coding solutions available in the literature that will be later used for subjective and objective quality assessment. As mentioned before, only the geometry component will be addressed. Naturally, the MPEG G-PCC and V-PCC codecs, currently under development, are the most relevant for this work as they are the most recent and efficient PC coding solutions available. These codecs are part of the MPEG-I set of standards, which aim to design key technologies for immersive media.

Considering the above context, the PCL, MPEG G-PCC and V-PCC codecs were selected. These codecs represent the three most relevant ways to structure PC data for coding purposes, namely tree, surface and patch, respectively. A tree is a data structure where the points are organized in a tree, e.g. octrees, kd-trees; a surface is a data structure where the points are represented with a parametrized surface model (e.g. represented as a set of triangles); finally, a patch clusters points into groups with some size, which is suitable for 3D to 2D projections. Naturally, these PC codecs produce different types of geometry artifacts, such as loss of geometric detail, geometric deformations, holes creation and other geometric distortions, e.g. curved surfaces represented by a set of planes.

#### 1) PC Coding with Tree Structures

The PC coding solution selected for this class is the popular PC codec public available in the Point Cloud Library [36], a large scale, open project for 2D/3D image and PC processing. To facilitate the compression of geometry data, this codec represents the PC 3D coordinates and its attributes with an octree structure [37]. The PCL PC codec is often used as benchmark since it can handle unorganized PCs of arbitrary size/density acquired with many types of sensors and has low encoding and decoding complexity.

In PCL, each octree node corresponds to a voxel in 3D space. The root node corresponds to a voxel that contains all points of the PC, the so-called *PC bounding box*. Then, starting from the root node, each voxel is divided iteratively into 8 voxels with the same size; naturally, a node is not divided if the corresponding voxel is not occupied. The occupancy of a node is represented with a single byte that signals the occupied child nodes up to the leaf voxels. By traversing the octree in breadth-first order, a stream of occupancy bytes is created, thus allowing an efficient representation of the PC geometry.

The decoded quality is determined by the octree depth, which indirectly specifies the minimum voxel size; this corresponds to a pruned octree, since the octree will not have the full depth. When the PC is decoded, all the points inside an occupied voxel are represented with just one point at the voxel center. The statistics of the occupancy bytes are exploited by an entropy encoder (range coder [38]) that takes into account the specific (non-uniform) symbol frequencies. The PCL v.1.8 version was used as the reference software for the experiments reported here. In these experiments, no point detail coding is performed to refine the geometry within the leaf voxels.

#### 2) PC Coding with Surface Models

The PC coding solution selected for this class is the MPEG G-PCC codec, which is capable of lossy and lossless coding of large PCs, with spatial random access, view dependent processing, packetization, and scalability [39]. As the PCL octree-based codec, the G-PCC codec is also based on octree decomposition to code the PC geometry but extends this coding paradigm with a parameterized surface model. As in PCL, a pruned octree is used but the geometry of the points at each leaf voxel is not represented by the voxel center; instead, a set of triangles is used to represent a surface formed by these points.

In G-PCC, the input PC data is first voxelized such that the resulting coordinates lie in the cube $[0, 2^d - 1]^3$ and all points are represented by the voxel center; $d$ corresponds to the octree (full) depth parameter (defined *a priori*). Then, a pruned octree is created, from the root down to some specific octree level ($\ell$), which must be smaller or equal than the octree depth; for lossless coding, level must be equal to the octree depth. If $\ell$ is smaller than depth, a polygonal representation is used to represent the points, which is known as *TriSoup*, an amalgam for Triangle Soup. This means that the limited depth octree is complemented with additional geometry information within groups of voxels, called *blocks*; this additional geometry is represented by *vertices*, corresponding to the intersections of the surface with some edges of the block (in this case at most 12 vertices). This set of vertices is sufficient to reconstruct a surface, corresponding to a non-planar polygon passing through the vertices. The test model category 1 (TMC1) v1.1 version of the G-PCC reference software was used for the experiments reported in this paper.

#### 3) PC Coding with Patch-based Projection

The PC coding solution selected for this class is the MPEG V-PCC codec, which targets dynamic PC coding and performs a 3D to 2D mapping of both the geometry and color components [40]. Thus, depth and texture images are created and can be coded with any video codec, notably a High Efficiency Video Coding (HEVC) standard-compliant codec [41].

In the first step, the PC is decomposed into several patches with smooth boundaries, while minimizing the reconstruction error. PC points are clustered according to the relation between their normals and the normal directions of six predefined oriented planes (forming a 3D bounding box). Then, patches are extracted from these clusters using a connected component technique and mapped onto a 2D grid using a packing process which attempts to minimize the unused space. Each $n \times n$ (e.g., 16×16) block in the grid is associated with a unique patch. After the packing, geometry (depth) and texture maps are created and the empty spaces between patches are filled using a padding process to obtain a smooth image (easier to code). These maps are passed to an HEVC encoder, which exploits the spatial and temporal correlations in a very efficient way. An occupancy map used to determine whether a grid cell is occupied or not is also coded to determine which 3D points are decoded. The V-PCC reference software used for the experiments reported in this paper was TMC2 v.2.

*C. Coding Artifacts*

This section describes the distortions associated with each of the PCC selected solutions. A characterization of the artifacts is important to understand the perceptual impact in the subjective tests and the limitations of the available objective metrics. For this purpose, some frames are extracted from the videos created for the subjective test session described in Section IV.B. All PCs were coded using the test conditions for low rate as described in Section IV.A. The selected PCs examples try to show as much as possible the most typical visual artifacts found during this study. A more complete set of examples, including these PCs for all codecs and rendering combinations (low quality only) are available in Section I of the supplementary material.

*1) PCL Codec*

In the PCL codec, as the target bitrate decreases (lower octree depth), the number of decoded points also decreases since all points inside a voxel are represented by just one point at the voxel center. The consequence is an increase of the distance between decoded points and thus lack of detail. When PCL decoded PCs are rendered, in any rendering solution, the lack of detail (i.e. points) results into a pixelated (or overly sub-sampled) decoded PC. An example of the artifacts produced with PCL coding at low rate is illustrated in Fig. 2, for the *Loot* PC for all rendering solutions. As shown, PCs are rather pixelated (*RPoint* and *RColor*) or lack detail (e.g. face and hands in *RMesh*).

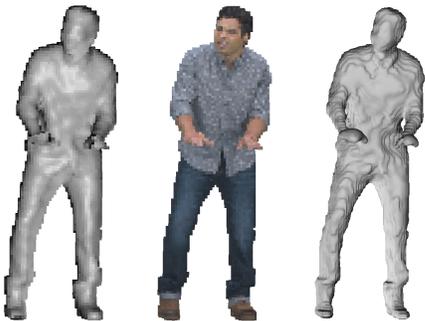

Fig. 2. PCL coding artifacts for *Loot* when rendering with *RPoint*, *RColor* and *RMesh* (from left to right).

*2) MPEG G-PCC Codec*

The MPEG G-PCC codec prunes the octree at some specific depth and after creates a surface representing all points in that depth with more precision. The rendering artifacts produced are very different from PCL, since the number of decoded points is no longer reduced. An example of the artifacts produced by G-PCC at low rate is illustrated in Fig. 3 for the *Egyptian Mask* PC for all rendering solutions. The geometric artifacts essentially come from the *TriSoup* process which may create false edges at the boundaries of the blocks or triangles; for low rates, these triangles may be visible. Moreover, when the PC is sparse in some region, the *TriSoup* process may cause artificial holes (with polygonal shapes) or even an increase in the size of holes already present in the original PC.

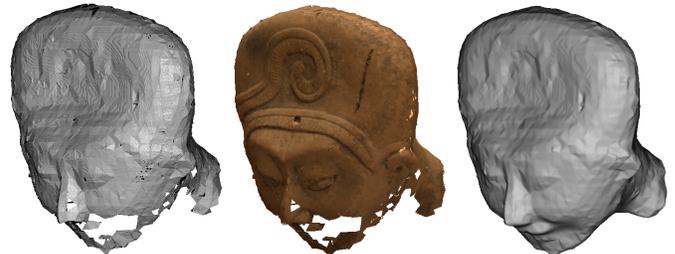

Fig. 3. G-PCC coding artifacts for *Egyptian Mask* when rendering with *RPoint*, *RColor* and *RMesh* (from left to right).

*3) MPEG V-PCC Codec*

In MPEG V-PCC, PC data is coded with traditional prediction and 2D transform tools. The more visible rendering artifacts correspond to blockiness and the creation of false edges, often associated to the directional Intra prediction modes. An example of the rendering artifacts produced by V-PCC is illustrated in Fig. 4 for *House without a Roof* PC for all rendering solutions. While false edges are visible, mostly for *RPoint* rendering, V-PCC distortions are not very visible for *RColor* rendering. For *RMesh*, the entire decoded PC is smoother compared to *RPoint*. However, some details are lost (e.g. in the bell tower), which may cause lower perceived quality.

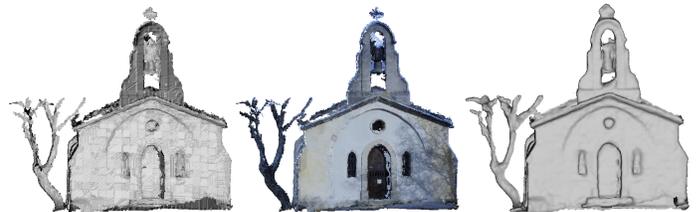

Fig. 4. V-PCC coding artifacts for *House without a Roof* when rendering with *RPoint*, *RColor* and *RMesh* (from left to right).

Another type of artifact typically occurring with V-PCC coding (not shown in this example) is the lack of points at the patch boundaries.

## IV. PC RENDERING AFTER CODING: SUBJECTIVE QUALITY ASSESSMENT STUDY

In this section, the creation of the visual content for the subjective experiments is described along with the test setup and the experimental results. The analysis of these results will allow to assess the visibility of distortions of each PC codec under different rendering approaches.

*A. Test Conditions*

Six static PCs have been selected from the MPEG content repository [42], notably *Egyptian Mask* and *Frog* from the class inanimate objects, *Facade9* and *House without a Roof* from the class buildings and facades, and *Longdress* and *Loot* from the class people. This selection includes PCs with different levels of coding complexity (as defined by MPEG in [43]), with four PCs from class A (lowest complexity) and two PCs from class B (medium complexity). These six selected PCs have rather different characteristics in terms of content type, geometry and color. Thus, the most important factors in the PCs selection were: i) the PC density (PCs sparse and dense), ii) the semantic type of content (PCs from Inanimate Objects, Facades & Buildings and People), iii) the PC geometry structure (PCs with holes and PCs with flat surfaces), and iv) the color characteristics (PCs with a small or high color gamut). Table I shows the PC name, number of points, coordinates precision, category while Fig. 5 shows the original PCs with *RColor* rendering.

TABLE I
TEST MATERIALS AND CHARACTERISTICS.

| PC Name | No. Points | Precision | Category |
|---|---|---|---|
| *Egyptian Mask* | 272,684 | 12 bit | Inanimate Objects |
| *Facade9* | 1,596,085 | 12 bit | Facades & Buildings |
| *Frog* | 3,614,251 | 12 bit | Inanimate Objects |
| *House wo. roof* | 4,848,745 | 12 bit | Facades & Buildings |
| *Loot* | 805,285 | 10 bit | People |
| *Longdress* | 857,966 | 10 bit | People |

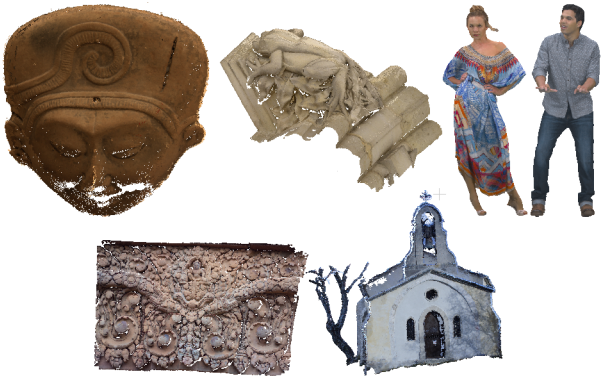

Fig. 5. Test materials with *RColor* rendering. From left to right and top to bottom: *Egyptian Mask, Frog, Longdress, Loot, Facade9* and *House without a Roof*.

The selected PCs were coded with the three selected PC codecs, at three different rates, to obtain decoded PCs with three different perceptual qualities, labelled as Low (L), Medium (M) and High (H). The selected codecs represent three different coding paradigms, notably PCL for tree structures, MPEG G-PCC for surface models and MPEG V-PCC for projection-based coding. For each of the MPEG PC codecs, three different rate points have been selected based on the suggested coding parameters in the MPEG Common Test Conditions (CTC) [43] for lossy coding. These rate points resulted into three distinguishable qualities, ranging from low to high. For PCL, the octree depth parameter was defined in a way to obtain a similar range of qualities compared to V-PCC. Table II shows the coding parameters used for the PCL and MPEG G-PCC codecs.

TABLE II
OCTREE DEPTH AND LEVEL FOR PCL AND G-PCC CODECS FOR LOW (L), MEDIUM (M) AND HIGH (H) QUALITIES.

| PC | PCL | | | G-PCC | | | |
|---|---|---|---|---|---|---|---|
| | Octree depth | | | Octree depth | Octree Level | | |
| | L | M | H | L/M/H | L | M | H |
| *Egyptian Mask* | 7 | 8 | 9 | 9 | 5 | 6 | 7 |
| *Frog* | 8 | 9 | 10 | 11 | 7 | 8 | 9 |
| *House wo. Roof* | 8 | 9 | 10 | 11 | 7 | 8 | 9 |
| *Facade9* | 8 | 9 | 10 | 11 | 7 | 8 | 9 |
| *Loot* | 7 | 8 | 9 | 10 | 6 | 7 | 8 |
| *Longdress* | 7 | 8 | 9 | 10 | 6 | 7 | 8 |

For G-PCC, the octree depth establishes the PC precision (after the encoding voxelization step). The *level* parameter corresponds to some octree layer after which a polygonal representation is used. For PCL, the octree depth (*OD*) is set indirectly, using the PCL Octree Resolution (*OR*) parameter which corresponds to the size of the voxel computed as $OR = 2^{(P-OD)}$, with $P$ as PC precision (defined in Table I). Table III shows the MPEG V-PCC HEVC quantization parameter (*QP*) used for depth map coding (note that no color coding is performed) and *B0* is the occupancy map precision. For V-PCC, all the test material was voxelized to 10-bit precision.

TABLE III
QUANTIZATION PARAMETER (QP) AND OCCUPANCY MAP PRECISION (B0) FOR V-PCC CODEC FOR LOW, MEDIUM AND HIGH QUALITIES.

| Quality | Low | Medium | High |
|---|---|---|---|
| QP | 32 | 24 | 16 |
| B0 | 4 | 4 | 4 |

B. *Test Sessions*

The subjective quality assessment was performed in three test sessions, each using a different PC rendering approach. Following Section III.A, the test sessions have been labelled as:

1. **RPoint** session: PCs are rendered with point-based rendering with point shading without color attributes.
2. **RColor** session: PCs are rendered with point-based rendering with the original color (by recoloring) and no shading.
3. **RMesh** session: PCs are rendered with mesh-based rendering with surface shading without color attributes.

The PCs were visualized in a non-interactive way, which means that the original and decoded PCs were rendered to standard video sequences and shown on a 2D display. The advantage of such approach is that all subjects in the subjective test see the same parts of the PC exactly in the same way, thus obtaining more reliable subjective assessment scores. The *CloudCompare* PC processing software was used for rendering with the point size, normal estimation and surface reconstruction performed as described in Section III.A. The lighting conditions, which influence the shading process in *RPoint* and *RMesh,* correspond to the default conditions, this means ambient light source (sun light) and no spotlight. A simple camera path rotation around the object was used to create a 2D rendered video; this path was found to allow a complete visualization of the PC and, most importantly, the coding artifacts under evaluation. For some PCs (e.g. *Facade9*), no geometry was acquired for the back side and, thus, the rotation path was restricted to the frontal part of the object. The virtual distance between the PC and the camera did not change, similarly to standard image and video subjective test methodologies where the distance between the subject and the display is fixed.

The rendered videos have a spatial resolution of 1600×800, a temporal resolution of 25 frames per second (fps), and a duration of 10 seconds. For all the three sessions, the rendered videos were visualized on a 23-inch ASUS VH238 monitor with 1920×1080 resolution. An i7 workstation with the Intel HD 530 graphic card and 128MB video memory was used to play the rendered videos at the correct frame rate.

*C. Subjective Quality Assessment Methodology*

The PCs selected for the subjective study have rather different characteristics. Due to the acquisition process, some original PCs can be rather noisy, e.g. MPEG cultural heritage and buildings sub-category may have holes, outliers or even positioning errors. Also, the density (number of points per unit volume) of the original PC may have a significant impact on the perceived quality of the original rendered PC. These two factors may affect the subjective scores given by the subjects. Since these issues affect both the original and decoded PCs, the DSIS subjective test methodology was selected for all the test sessions of this subjective study. Thus, subjects visualize first the original and then the decoded rendered PCs and score the similarity of decoded PC relatively to the original, which allows to mitigate the impact of acquisition artifacts and other original PC characteristics.

There were 20 subjects participating in each test session with 18 people participating in all the three sessions and four people in one or two sessions. At the beginning of each session, the goal of the subjective assessment experiment was explained to the subjects and they were asked to participate in a short training session to become familiar with the application interface. For the training sessions, the *Statue Klimt* PC from the same MPEG repository was used.

The full set of rendered PCs was organized into six rounds per session with each round including all PCs with one of the three levels of quality. Since there were six PCs coded with three different codecs for three rate points, $6 \times 3 \times 3 = 54$ stimuli were assessed in each session. According to Recommendation BT-500.13 [44], the subjects see first the original rendered PC and after the impaired (this means decoded) rendered PC and score the later in a 1-5 scale associated to five impairment levels, notably very annoying, annoying, slightly annoying, perceptible but not annoying and imperceptible. The display of each new rendered video was controlled by the subjects by pressing 'Play'. The subjects had the option to replay both video sequences (original followed by impaired PCs) before giving the subjective score. This option allows to reduce the cognitive load of the subjects and, thus, obtain more reliable scores. Each session had a duration of approximately 28 minutes, considering the training and scoring times. To avoid that the results of one session influenced the results of another session, a minimum of 48h between test sessions was respected.

For each session, outlier subjects were identified based on the collected scores, following the procedure in BT.500-13 [44]; only one outlier was identified in the *RMesh* session. After, the average of all scores across the subjects were computed for each test PC, thus obtaining a MOS for each PC under evaluation. The subjective scores for the three test sessions along with the original and decoded rendered PCs are publicly available at [45] and, thus, may be used by the research community.

*D. Experimental Results and Analysis*

The focus of this section is on the study of the impact of different PC rendering solutions on the user perceived quality for PCs compressed with different coding artifacts. Additionally, the obtained subjective scores are analyzed to assess the visibility of the different coding artifacts. The subjective scores obtained for the three test sessions will be the basis for this study; in this case, the MOS values represent the similarity between the original and decoded PCs and not the intrinsic PC quality for which many other factors play a role. From now on, when the 'quality' term is used, it regards only to the fidelity (or similarity) aspect.

*1) Impact of Rendering on Perceived PC Quality*

This section studies the impact of the three rendering solutions on the perceived PC quality. Note that, within each session, the rendering methods were not mixed and thus the subjects evaluated videos from decoded PCs for each rendering solution independently.

Fig. 6 shows the 54 MOS for all PCs within each test session (each associated to a rendering solution). In Fig. 6**,** the MOS are sorted in ascending order, thus from lower to higher scores; each score is labelled with a rendered PC index and corresponds to a coding condition. To identify which are the most frequent MOS per session (data not shown in Fig. 6**),** Fig. 7 shows the MOS distributions (number of votes) given by the subjects in the three rendering sessions.

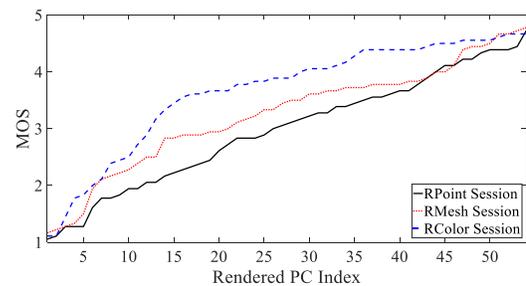
Fig. 6. Sorted MOS for all test PCs in the three test sessions.

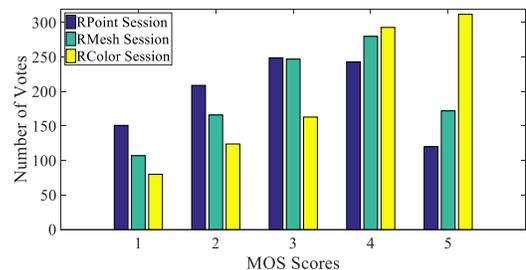
Fig. 7. MOS histograms for the three test/rendering sessions.

Fig. 6 shows that the scores are well distributed over the full range, from low (close to 1) to high (close to 5). The *RColor* session (blue curve) shows the highest MOS, followed by the *RMesh* session and, finally, the *RPoint* session.

**RPoint rendering:** The geometry coding distortions are more visible for *RPoint* rendering since *RMesh* and *RColor* have mechanisms to mitigate the visual impact of the coding artifacts, e.g. filtering or masking. This can be clearly observed in Fig. 6 where the coding artifacts are more visible for the curve with lower MOS and, thus, as shown in Fig. 7, more '1', '2' and '3' votes are obtained for *RPoint* compared to *RMesh* and *RColor*.

**RMesh rendering:** As shown in Fig. 7, *RMesh* rendering has higher MOS (and less low MOS) than *RPoint* rendering. This can be explained by the fact that *RMesh* rendering includes a surface reconstruction process (polygonal mesh creation) which smooths the PC and makes the coding distortions less visible, somehow behaving as a denoising filter. However, it should be emphasized that PC edges and details are also smoothed with this type of rendering and, thus, *RMesh* is not able to outperform a point-based rendering solution with color (*RColor*), where the points are simply rendered with a basic primitive, e.g. circles or squares. It also requires the extra pre-processing step of surface reconstruction before rendering, which may be difficult to apply in some application scenarios due to the scene complexity or the PC size (number of points).

*RColor* **rendering:** For *RColor* rendering, the original texture contains natural shading information, acquired from the light reflected by the object surface. This contrasts with the *RPoint* and *RMesh* renderings, which use a single color with synthetic shading and the final result depends on the accuracy of the (extracted) normal vectors (geometry only). However, it is clear that the original color captured during the acquisition is able to mask many of the geometric distortions, changing the perceived surface of the objects. Also, the texture details are able to hide geometric distortions since the human visual system is less sensible to high frequencies; this will cause the subjects to perceive less distorted shapes and, thus, better scores are given in the *RColor* session. In fact, in the *RColor* session, most of the scores are '4' and '5' as shown in Fig. 7, which means that most of the decoded PCs were considered to have high similarity with the original PCs. In this case, the most visible geometric distortions are limited to the object boundaries.

In summary, rendering with high quality color attributes masks the geometric distortions and results in higher perceived PC qualities. However, color attributes may not be available, and some applications may require high geometry fidelity. For example, geographical information systems and cultural heritage applications typically only tolerate imperceptible geometry deformations; in such cases, *RPoint* rendering could be an appropriate choice to avoid the influence of color masking or geometry filtering. On the other hand, if color is not available and geometry degradations are tolerable if not visible, *RMesh* rendering should be used, since it allows to mitigate the impact of some coding artifacts, e.g. holes or false edges, compared to *RPoint*, thus leading to higher perceived PC qualities.

*2) Impact of Rendering on the Coding Artifacts Visibility*

This section studies the impact of the three rendering solutions on the visibility of the coding artifacts associated to the three selected codecs. With this purpose in mind, the MOS for the three PC codecs and the three rendering solutions are shown in Fig. 8.

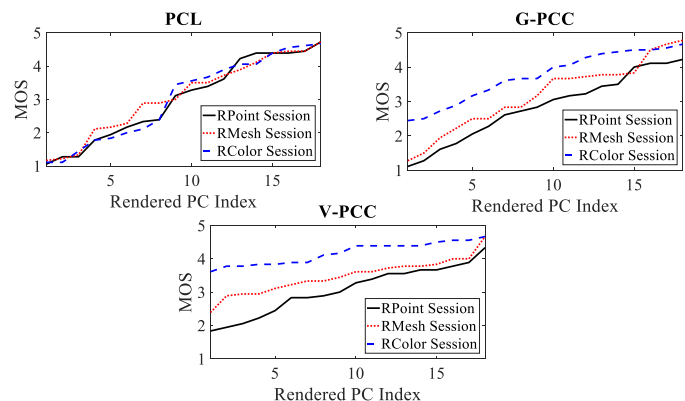

Fig. 8. MOS for each PC codec, organized by rendering approach.

This PC codec presentation of the MOS, more granular compared to Fig. 6, allows comparing the impact of the rendering solution on the final perceived visibility, when different coding artifacts are present. From Fig. 8, it is clear that the MOS distribution for each rendering approach is not similar for all codecs. The main conclusion is that the different types of coding artifacts are not equally visible for all rendering approaches. Based on the results, the following conclusions about the sensibility of the various rendering solutions to the PC codecs considered, and thus type of coding artifacts, may be derived:

**PCL Coding:** PCL distortions are visible regardless of the rendering solution and, thus, MOS are rather well distributed in the 1-5 range. This is mainly because a pure octree PC coding solution controls the decoded quality by limiting its maximum depth and, thus, decoded PCs have a lower number of points than the original PCs, sometimes significantly lower. Thus, larger point sizes for *RPoint* and *RColor* rendering are needed, creating a pixelated effect (perceptually unpleasant). Although a surface is reconstructed with *RMesh* rendering, when the number of points is reduced, details are lost and some meshes even show geometric distortions due to the surface reconstruction process.

**G-PCC Coding:** G-PCC distortions are less visible for *RColor* rendering compared to *RPoint* and *RMesh*, since the color masks the surface distortions. However, false edges, holes and geometric distortions at boundaries are still visible for severe distortion cases. *RPoint* and *RMesh* follow a similar trend, with slight better scores for *RMesh*, since it mitigates the impact of some coding artifacts (e.g. holes or false edges), thus offering a more visually appealing surface.

**V-PCC Coding:** V-PCC distortions are not very visible for *RColor* rendering since they are not large enough to create strong deformations and the color masks most of the geometric distortions. Due to the V-PCC projection onto 2D maps (texture and depth) and the efficient HEVC coding process, most of the surfaces are consistently represented, although with some error regarding the original surface. V-PCC distortions are also less visible for *RMesh* than for *RPoint* rendering due to the impact of surface reconstruction-based rendering on the perceived quality.

*3) Statistical Significance Analysis of Subjective Assessment*

This section presents a statistical significance analysis of the subjective quality assessment. The goal is to evaluate if the differences between the MOS for the three rendering approaches (*RPoint*, *RColor* and *RMesh*) are statistically significant at a given confidence level. Base on procedures suggested in previous work [46][47][48], three statistical tests were applied. For all tests, the full set of obtained scores was divided in three groups, one group of scores for each PC rendering approach, since the results being tested for statistical significance (sections IV.D.1 and IV.D.2) evaluate the impact of rendering. The selection of these tests was motivated by the fact that for V-PCC data the variance homogeneity test fails according to a Levene's test while the distribution of data is not normal for the All case according to the Shapiro-Wilk normality test.

*Welch ANOVA significance test:* To evaluate if the dependency of MOS values on the rendering method is statistically significant, the Welch ANOVA significance test was applied, thus comparing groups of MOS values, one group for each rendering method. This test measures the difference between the mean values of each group with a 5% significance level without requiring homogeneity of variances. The null hypothesis assumes that MOS values for the various groups (rendering methods) are drawn from a population with equal means. Table IV shows the p-values and associated MOS averages when considering all possible groups of scores for each codec ('PCL', 'G-PCC' and 'V-PCC' columns) and for all the codecs together ('ALL' column). When the p-value is lower than 0.05 (significance level), the separation between these rendering approaches is statistically significant.

TABLE IV.
P-VALUES FOR THE WELCH ANOVA SIGNIFICANCE TEST AND MOS AVERAGES FOR EACH SESSION, I.E. *RPOINT*, *RCOLOR*, *RMESH*.

|  | PCL | G-PCC | V-PCC | All |
|---|---|---|---|---|
| p-value | 0.98 | 0.019 | 0.0 | 0.002 |
| MOS averages | 3.0, 3.0, 3.1 | 2.8. 3.8, 3.2 | 3.1, 4.2, 3.5 | 3.0, 3.7, 3.2 |

***Games-Howell post-hoc and Wilcoxon signed-rank tests:*** To compare the several possible pairs of rendering methods (i.e. perform a multiple-comparison statistical test), the Games-Howell post-hoc test was selected since it also does not require the homogeneity of variances, again with a 5% significance level. Table V shows the p-values obtained for this post-hoc test for all possible rendering pairs. Moreover, since MOS values obtained do not follow a normal distribution (i.e. normality does not hold) for the 'ALL' case, the p-values obtained for the Wilcoxon signed-rank test (5% significance level) are shown in Table VI. The Wilcoxon signed-rank test assesses whether the group mean ranks differ; as it is non-parametric, i.e. it does not assume any data distribution, it is thus more suitable for this case. For the Games-Howell post-hoc and Wilcoxon signed-rank tests, when the p-value is lower than 0.05 (significance level), there is statistical significance between groups of MOS.

TABLE V.
P-VALUES FOR THE GAMES-HOWELL POST-HOC TEST FOR ALL RENDERING PAIRS (PAIR ORDER IS IRRELEVANT).

|  | PCL | G-PCC | V-PCC | All |
|---|---|---|---|---|
| *RPoint↔RColor* | 0.998 | 0.011 | 0.000 | 0.002 |
| *RPoint↔RMesh* | 0.973 | 0.586 | 0.144 | 0.300 |
| *RColor↔RMesh* | 0.988 | 0.162 | 0.000 | 0.085 |

TABLE VI.
P-VALUES FOR THE WILCOXON TEST FOR THE 'ALL' CASE.

|  | *RPoint↔RColor* | *RPoint↔RMesh* | *RColor↔RMesh* |
|---|---|---|---|
| p-value | 0.0 | 0.023 | 0.003 |

***Final remarks:*** From the analysis of the results in Table IV and Table V above, i.e. Welch ANOVA significance and Games-Howell post-hoc tests, respectively, the analysis of section IV.D.2 can be confirmed and new conclusions can be derived:

PCL: The difference between the MOS for the three rendering approaches is not statistically significant and thus, any rendering can be used. This was expected since PCL distortions are visible regardless of the rendering solutions and, thus, similar subjective scores were obtained for all rendering approaches.

G-PCC: *RColor* is better than *RPoint* and, thus, if color is available, it should be used in point-based rendering solutions. There is no statistical difference between *RPoint* and *RMesh* and

*RColor* and *RMesh*, meaning that there is no advantage in using mesh-based rendering (which may even require complex surface reconstruction methods).

V-PCC: RColor is better than *RPoint* and *RMesh* and, thus, color effectively masks the geometric distortions associated to the V-PCC coding artifacts. For the 2$^{nd}$ best rendering method, there is no statistical difference between *RPoint* and *RMesh* and thus, this means that any of these two methods could be used.

Finally, from the analysis of the results in Table VI above, i.e. Wilcoxon signed-rank test results, statistical significance was obtained for all rendering pairs (i.e. *RPoint↔RColor*, *RPoint↔RMesh*, and *RColor↔RMesh*) for the 'ALL' case, meaning that a ranking order of the rendering methods is established. The results for this test show that *RColor* is statistically better than *RMesh* and *RMesh* is statistically better than *RPoint*. This confirms the intuitive ordering shown in Fig. 6 and the conclusions in section IV.D.1.

V. PC RENDERING AFTER CODING: OBJECTIVE METRICS PERFORMANCE ASSESSMENT

The main purpose of this section is to evaluate the performance of several PC objective metrics in the presence of coding artifacts. Only metrics accounting for geometry errors are considered since this is the PC component where artifacts may cause a higher negative impact on the user quality of experience.

First, the objective metrics that are evaluated are presented and then the correlation between the objective metrics and the MOS are reported and analyzed. This will allow to understand which objective metrics perform better, which type of coding degradations can be more appropriately accounted and the impact of the rendering solution on the PC objective metrics accuracy.

*A. Geometry Objective Assessment Metrics*

Objective assessment metrics for PCs are essential for several tasks, notably: i) measuring the PC fidelity and thus playing a part on the rate-distortion (RD) performance assessment of PC coding solutions; ii) optimizing PC coding solutions, e.g. allowing to make perceptual optimizations; iii) measuring end-to-end quality in PC streaming solutions, thus involving other degradations besides coding.

In this section, the most popular geometry objective metrics for PC quality assessment adopted for this study are presented. These quality metrics are full reference metrics and measure the level of similarity (impairment level) of the decoded PC (with some coding artifacts) with respect to the original PC. Overall, they are based on establishing correspondences and computing distances between points of the original and decoded PCs. The following classes of metrics can be defined:

1. **Point-to-point (Po2Point):** Score depends on the distance between corresponding points.
2. **Point-to-plane (Po2Plane):** Score depends on the distance between a point and a reference plane where this plane is a representation of the surface around a point in the original PC.
3. **Plane-to-plane (Pl2Plane):** Score depends on the similarity of the planes representing the surfaces near corresponding points.

While Po2Point metrics are rather straightforward, Po2Plane metrics use a plane to represent the surface around a point, under the assumption that this is a reasonable representation of the object surface in a specific region. The Pl2Plane metrics extend this concept and use two planes to represent the surfaces around points, both in the original and degraded PCs.

*1) Point-to-Point (Po2Point) Objective Metrics*

Point-to-point objective metrics establish point-wise correspondences in two directions: 1) direct $R \to T$: for each point in the original (as reference) PC R and the nearest neighbor (NN) point in the degraded (as test) PC T [49]; 2) inverse $T \to R$: correspondences are computed similarly to 1) but in the opposite direction, thus, from PC T to PC R.

Assuming $\vec{e}_1(r_i, t_j)$ as an error vector between point $r_i$ in PC R and the $r_i$ nearest neighbor point $t_j$ in PC T, the Po2Point error vector length, i.e. the distance $d_{R,T}^{Po2Point}$ between these two points is given by:

$$d_{R,T}^{Po2Point}(i) = \|\vec{e}_1(r_i, t_j)\|_2^2 \qquad (1)$$

This distance is computed for all the points in both directions, i.e. from original to degraded PCs, $d_{R,T}^{Po2Point}$, and from degraded to original PCs, $d_{T,R}^{Po2Point}$, for every point. There are three main approaches to aggregate or pool all the computed errors:

- **Mean Squared Error (MSE):** Average of the squared distance between each point and their corresponding nearest neighbor, for all points, as defined in:

$$\text{MSE}_{R,T} = \frac{1}{N_R} \sum_{\forall r_i \in R} d_{R,T}^{Po2Point}(i) \qquad (2)$$

where $N_R$ is the number of points in the original PC R.

- **Hausdorff (HAUS) distance:** Maximum for all points of the MSE distance as defined in:

$$\text{HAUS}_{R,T} = \max_{r_i \in R} \{d_{R,T}^{Po2Point}(i)\} \qquad (3)$$

- **Geometric PSNR:** Geometric PSNR metric defined as:

$$\text{PSNR}_{R,T} = 10 \log_{10}\left(\frac{3P^2}{\text{MSE}_{R,T}}\right) \text{ with } P = 2^{pr} - 1 \qquad (4)$$

where $P$ is the peak constant value and $pr$ the PC coordinates precision. The metrics above defined just for one direction (from PC R to T) are computed also in the other direction and, thus, the final metric value can be computed as:

$$\text{MSE} = \max(\text{MSE}_{R,T}, \text{MSE}_{T,R}) \qquad (5)$$

$$\text{HAUS} = \max(\text{HAUS}_{R,T}, \text{HAUS}_{T,R}) \qquad (6)$$

$$\text{PSNR} = \min(\text{PSNR}_{R,T}, \text{PSNR}_{T,R}) \qquad (7)$$

The Po2Point PSNR metric is used nowadays by the MPEG 3D Graphics Compression (3DGC) group in the evaluation of PC coding methods such as G-PCC and V-PCC, labelled as D1 [42].

*2) Point-to-Plane (Po2Plane) Objective Metrics*

Point-to-plane metrics take into consideration the underlying object surface represented with the PC by fitting a plane to the local neighborhood of each point [50]. Considering the 3D point locations and their associated surfaces, the normal for each point is equal to the normal of the tangent plane to the surface. A point and the corresponding normal vector can, thus, determine the tangent plane for each point. As for Po2Point metrics, Po2Plane metrics are also symmetrically computed for both directions, i.e. from original to degraded and from degraded to original PCs. However, Po2Plane metrics require the computation of normals on the original PC, which are directly used in the direct direction $(R \to T)$. For the opposite direction $(T \to R)$, the normal for each point is estimated by averaging the normals of the nearest neighbor points from the original PC.

The Po2Plane error distance between two points $\vec{e}_2(r_i, t_j)$ is obtained by first computing the Po2Point error vector $\vec{e}_1$ which is then projected onto the normal $\vec{n}_j^t$. Thus, the Po2Plane distance $d_{R,T}^{Po2Plane}(i)$ that represents the error between a point and its corresponding surface is given by:

$$d_{R,T}^{Po2Plane}(i) = \|\vec{e}_2(r_i, t_j)\|_2^2 = (\vec{e}_1(r_i, t_j) \cdot \vec{n}_j^t)^2 \qquad (8)$$

MSE distance, Hausdorff distance and Geometric PSNR may then be computed with the projected error distances as for Po2Point metrics (where the error vector is not projected). In this way, the degraded PC points that are closer to the reference surface have smaller projected distances even though they are farther from the corresponding point on the original PC. The Po2Plane PSNR metric is also used by MPEG for the evaluation of the G-PCC and V-PCC codecs, labelled as D2 [42].

*3) Plane-to-Plane (Pl2Plane) Objective Metrics*

This type of objective metrics estimates the similarity of surfaces in the original and degraded PCs [51]. In this case, tangent planes are estimated for both the original and degraded points. As for Po2Plane metrics, tangent planes are used as a local linear approximation of the underlying object surface but, in this case, planes are estimated for both the original and degraded PCs.

Again, to compute Pl2Plane metrics, the nearest neighbor correspondences are computed in both directions. The Pl2Plane metrics depend on the angular similarity (or dissimilarity) between the planes, i.e. the angular difference between local planes associated to the points in a correspondence. In this case, the so-called *cosine similarity measure*, $cs$, measuring the cosine of the angle between two vectors is used. The two vectors correspond to the normal vectors (perpendicular to the tangent planes) for the two points in a correspondence in PCs R and T [51], as in:

$$cs(i) = \cos(\theta_i) = \frac{\vec{n}_i^r \cdot \vec{n}_j^t}{\|\vec{n}_i^r\|_2 \|\vec{n}_j^t\|_2} \qquad (9)$$

where $\vec{n}_i^r$ and $\vec{n}_j^t$ are normals for points $r_i$ and $t_j$ in PCs R and T, respectively. To compute the angular difference (or distance), $d_{R,T}^{Pl2Plane}$, the inverse cosine is used as follows:

$$d_{R,T}^{Pl2Plane}(i) = 1 - \frac{2\arccos(|cs(i)|)}{\pi} \qquad (10)$$

After determining the angular difference for all the points in the original PC, different strategies for pooling, i.e. for aggregating the angular differences obtained for all points, can be defined. In this case, three pooling strategies were defined:

$$\text{MAD}_{R,T} = \frac{1}{N_R} \sum_{\forall r_i \in R} d_{R,T}^{Pl2Plane}(i) \qquad (11)$$

$$\text{MSAD}_{R,T} = \frac{1}{N_R} \sum_{\forall r_i \in R} \left( d_{R,T}^{\text{Pl2Plane}}(i) \right)^2 \quad (12)$$

$$\text{RMSAD}_{R,T} = \sqrt{\text{MSAD}_{R,T}} \quad (13)$$

MAD stands for mean angular difference, MSAD for mean squared angular difference and RMSAD as the square root of MSAD. As for the other types of metrics, (11)-(13) are performed symmetrically, this means in both direct and inverse directions, and the minimum value is selected as the final (similarity) score.

Since PCs have different precisions (depth) for the point coordinates, the error vectors for all these metrics may not be directly comparable. To overcome this problem, all PCs are normalized to have coordinates in the [0,1] range before computing the metrics. The only exception is the PSNR for Po2Point and Po2Plane metrics, which includes the peak $P$ that already plays the role of a scaling factor depending on the bit depth of each PC under evaluation.

### B. Experimental Results and Analysis

In this section, the selected objective metrics performance will be presented and analyzed for the subjective scores obtained in the three test sessions, thus for different rendering approaches. As recommended in [44][52], before assessing the objective metrics performance, a nonlinear logistic fitting has been applied on the objective scores to map them to the subjective scores scale. To assess the metrics performance, the *Pearson Linear Correlation Coefficient* (PLCC) is computed as a measure of the linear dependence between the MOS and each objective metric.

Table VII shows the PLCC for the 9 metrics described in the previous section, for each rendering approach, independently computed for each PC codec and also considering all codecs simultaneously (column All). With these results, the performance of each metric can be assessed for each of the three test sessions described in Section IV.B. A detailed analysis of the results in Table VII is presented in the following. First, from the perspective of the PC codec and coding distortions, after from the perspective of the rendering solution and, finally, assessing which metric performs the best and in which conditions.

#### 1) Impact of Coding on the PC Metric Assessment

**PCL Coding:** For PCL coded data, the Po2Plane and Po2Point metrics have the best PLCC (overall, PSNR is the best) with high correlations for all rendering approaches as shown in Table VII. As PCL controls the rate by reducing the number of decoded points, large objective errors and perceived distortions are visible for all sessions. This was expected since, when the compression ratio increases (lower rates), more and more points are discarded (due to octree pruning) and the remaining points are represented farther away from the original surface. PCL artifacts are strong enough to be visible even after the *RMesh* surface reconstruction.

**G-PCC & V-PCC Coding:** As shown in Table VII, the objective metrics performance for G-PCC is slightly lower (4 to 5%) compared to PCL and shows the highest performance for Po2Point metrics (only for *RPoint* and *RColor* sessions). Moreover, none of the selected objective metrics performs well for V-PCC coded data. The selected metrics underestimate the perceived similarity between original and degraded PCs, especially for *RPoint* and *RMesh* renderings where the geometric errors are less visible, e.g. compared to *RColor*. Since both the G-PCC and V-PCC codecs tend to add points with respect to the original PC (see Fig. 9), the density of points is increased and, thus, the perceived quality is higher (higher MOS). However, the objective metrics are not able to account for this effect and, thus, underperform for G-PCC and V-PCC codecs. In addition, since a wide range of values is obtained for the ratio of decoded over original number of points, notably depending on the codec (and also coding parameters), it is rather difficult to map errors to a perceptually meaningful metric; this makes the task of designing reliable objective metrics harder, especially when different types of codecs, with different coding artifacts, are jointly assessed ('All' column in Table VII). The correlation of the objective metrics for V-PCC is much lower compared to G-PCC (cf. Table VII). The projection-based V-PCC codec causes slight distortions on the geometry which are not very visible even for lower bitrates. On the other hand, G-PCC artifacts are more visible especially when the surface estimation (triangulation) process fails.

TABLE VII
PLCC (%) BETWEEN OBJECTIVE GEOMETRY ASSESSMENT METRICS AND MOS FOR THE THREE RENDERING APPROACHES. IN BOLD, THE BEST PLCC VALUE BETWEEN THE SUBJECTIVE AND OBJECTIVE SCORES AND ALL THE OTHER PLCC VALUES THAT DO NOT DEVIATE MORE THAN 0.02 FROM THE BEST PLCC.

| Type | Metric | RPoint | | | | RColor | | | | RMesh | | | |
|---|---|---|---|---|---|---|---|---|---|---|---|---|---|
| | | PCL | G-PCC | V-PCC | All | PCL | G-PCC | V-PCC | All | PCL | G-PCC | V-PCC | All |
| Point-to-Point | MSE | 84.5 | 53.7 | 26.3 | 51.9 | 84.1 | **85.5** | 44.1 | 64.3 | **90.5** | 32.7 | 7.5 | 39.0 |
| | HAUS | **90.5** | 45.6 | 34.2 | 23.7 | 87.1 | 59.2 | 57.8 | 18.6 | 88.3 | 49.4 | **31.2** | 32.5 |
| | PSNR | 87.4 | **86.5** | **55.0** | 66.9 | **89.8** | 71.1 | 72.3 | **78.3** | 91.6 | 51.7 | 18.3 | 68.8 |
| Point-to-Plane | MSE | 84.4 | 50.2 | 32.8 | 46.9 | 84.7 | 80.6 | 18.3 | 60.2 | 88.5 | 37.1 | 12.4 | 34.5 |
| | HAUS | **90.1** | 55.1 | 29.8 | 30.1 | 87.0 | 67.1 | 69.0 | 21.1 | 87.7 | 45.9 | 26.9 | 28.9 |
| | PSNR | **90.1** | 82.4 | 52.1 | **69.7** | 90.3 | 54.6 | 61.4 | **78.2** | 91.0 | 63.2 | 28.0 | **72.1** |
| Plane-to-Plane | MAD | 72.4 | 55.5 | 54.1 | 51.8 | 55.7 | 68.3 | **74.5** | 24.7 | 40.0 | 28.0 | 28.7 | 30.3 |
| | MSAD | 72.4 | 55.4 | 52.6 | 51.8 | 55.7 | 68.3 | **74.5** | 24.6 | 40.0 | 27.9 | 27.1 | 30.5 |
| | RMSAD | 71.7 | 55.6 | 51.8 | 51.5 | 55.0 | 68.0 | **72.9** | 24.7 | 39.8 | 26.2 | 30.4 | 31.1 |
| - | No. Points | 65.2 | 21.8 | 27.9 | 12.3 | 69.0 | 27.1 | 60.6 | 43.1 | 68.6 | 35.1 | 28.9 | 2.3 |

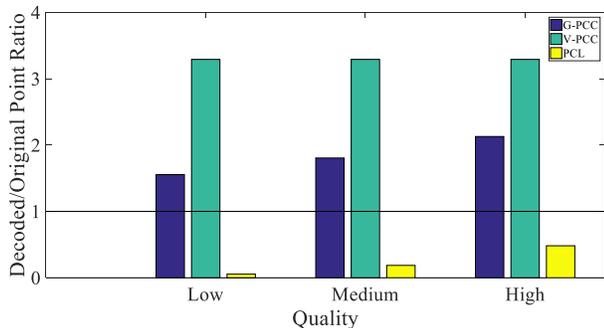

Fig. 9. Average ratio of decoded over the original number of points (1 means the original and decoded number of points are the same).

Fig. 10 shows the Po2Point RMSE distance errors histogram for all the points in two PCs coded with G-PCC and V-PCC for which the same overall Po2Point PSNR (60 dB) was obtained. As shown, although the PSNR is the same, the error distribution is very different between G-PCC and V-PCC. V-PCC errors are closer to zero which makes them less perceptually visible while G-PCC errors have larger magnitudes and, thus, are more visible. This implies that G-PCC has a lower subjective similarity (MOS of 1.1) than V-PCC (MOS of 3.15) even when the Po2Point PSNR objective metric computes the same DSIS score (in this case 60 dB). This observation happens also for other objective metrics, such as Po2Point and Po2Plane MSE.

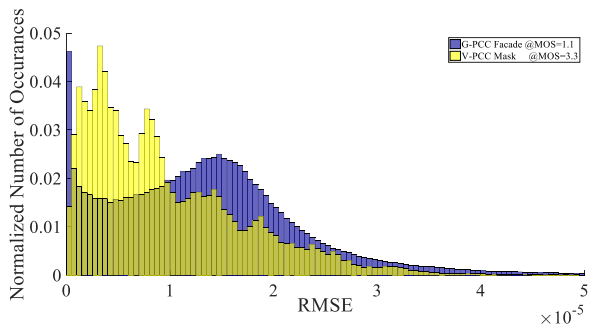

Fig. 10. Po2Point RMSE histograms for G-PCC and V-PCC.

In summary, the objective metrics performance in terms of correlation with MOS highly depends on the coding distortions introduced, being satisfactory when PCs are coded with PCL and G-PCC and performing poorly for the V-PCC codec. Naturally, no objective metric performs well for all codecs together, a real problem when comparing the RD performance of very different coding paradigms.

*2) Impact of Rendering on the PC Metrics Assessment*

As previously concluded, rendering can significantly influence the visibility of coding artifacts (i.e. the perceived similarity between original and decoded PCs) and, thus, it is important to analyze the objective metrics performance for different rendering solutions. The PLCC correlation over *all data* for all sessions is rather low (less than 78.4%) because the objective metrics cannot measure with accuracy all different types of distortions. However, as shown in Table VII, the best PLCC correlations occur for *RColor* rendering, for which higher MOS values were obtained. Thus, geometry metrics measure the perceived similarity better when color attributes are used.

The main reason is because subjects were able to better perceive degradations for medium and high quality ranges (which occur often with *RColor, cf.* Section IV.D) compared to low and medium quality ranges (which occur more often with *RPoint and RMesh, cf.* Section IV.D). For *RMesh* rendering, PLCC correlations are rather low comparing to the other rendering approaches, especially for G-PCC and V-PCC codecs. For *RMesh,* PC data is converted to a polygonal mesh (surface reconstruction) for rendering and most the objective metrics have low correlation performance for this type of representation.

In summary, objective metrics account better distortion artifacts and are more reliable when point-based rendering (with and without color, *RPoint* and *RColor*) is used to process the decoded PCs before visualization.

*3) PC Objective Metrics Correlation Assessment*

**Po2Point metrics:** Po2Point metrics have a high PLCC performance for many cases but are especially better than others for the PCL and G-PCC codec (*RColor* and *RPoint*). This is because PCL and G-PCC to some extent are an octree-based PC codec and, thus, some distortions still come from the positioning error related to the 3D partitioning of space into voxels, the target of this type of metrics. The Po2Point and Po2Plane Hausdorff metrics can also reach high PLCC performance, especially for PCL data and for the *RPoint* session (90.07). However, Hausdorff is not a very reliable metric when different types of coding distortions (all data and G-PCC/V-PCC) are considered together. The main reason is that only the maximum error is accounted and, thus, it is too sensible to outliers; this problem has been already observed in the literature [23] and can be mitigated using average pooling as in MSE and PSNR metrics.

**Po2Plane metrics:** Regarding Po2Plane metrics, the performance is very similar to Po2Point metrics, slightly better for some cases, since it considers the underlying surface from which the 3D point locations were sampled. Moreover, the Po2Plane PSNR metric excels, being rather reliable and consistent for many cases, outperforming the corresponding MSE metric. The main reason is that the peak used (computed from the geometry coordinate precision) to convert MSE to PSNR values acts as an important normalizer.

**Pl2Plane metrics:** Pl2Plane metrics have, in general, worst PLCC performance when compared to Po2Point and Po2Plane metrics. This is mainly because it is rather difficult to obtain reliable normals for the decoded PC, especially when some types of coding artifacts are present (e.g. holes) or when the decoded PC is rather sparse [51]. However, these metrics seem to be the best choice for the V-PCC codec (for *RColor* and *RMesh* renderings) where geometry errors mainly come from coding artifacts in the depth maps and, thus, are more consistent among different parts of the PC.

As a curiosity, the number of decoded points could also be used as an objective metric, see last line of Table VII. As expected, this metric performs very poorly, especially for V-PCC data where the number of decoded points is typically larger than the number of original points and critically depends on some coding parameters, e.g. *B0* for the occupancy maps.

*4) Statistical Significance Analysis of Objective Assessment*

Besides the usual PLCC correlation evaluation, the difference in the performance of one objective metric with respect to another was assessed for statistically significance, using the procedure suggested in [53]. For that purpose, the prediction residuals were first calculated by subtracting the subjective scores from the predicted subjective values, obtained by applying a nonlinear logistic fitting to the objective scores. These prediction residuals were obtained for every PC objective metric. Then, the one-tailed F-test was applied to the prediction residuals, to assess if the difference in performance (PLCC) between any two given PC objective metrics is statistically significant at some significance level. In general, the significance level should be set based on the sample size (cardinality) being evaluated for an increased power of the test (i.e. probability of rejecting the null hypothesis when it is not true) [54]; since the cardinality of the prediction residuals is 18 for a single PC codec (PCL, G-PCC and V-PCC), a 0.2 significance level was used [54]. The F-test assumes that the samples are normally distributed, and thus the kurtosis test was used to verify whether all prediction residuals followed a Gaussian distribution, which was the case for all the objective metrics, except the No. Points.

In this work, the F-test null hypothesis is that the prediction residuals for the two objective metrics being compared are obtained from normal distributions with the same variance, which means that the pair of objective metrics under evaluation is statistically similar. The alternative hypothesis is that the prediction residuals for the two objective metrics being compared are obtained from normal distributions with different variances, which means that the pair of objective metrics under evaluation are statistically different. By computing the ratio between the variances of the two prediction residuals, the test statistic, F, was obtained, which was then compared to the F-test critical value, Fcritical; the F-test critical value depends on the significance level and on the sample sizes. When F is higher than Fcritical, then the null hypothesis can be rejected, which means that the objective metrics under evaluation are statistically different; otherwise, the null hypothesis cannot be rejected, meaning that the objective metrics under evaluation are statistically indistinguishable. Since the test statistic F is always computed with the objective metric with larger prediction residual variance in the numerator, objective metric in the denominator corresponds to the metric with the best performance whenever the null hypothesis is rejected.

The statistical significance tables obtained are presented in Section II of the supplementary material. The results obtained confirm that P2Point and Po2Plane metrics have the best overall performance for many coding scenarios (PCL, G-PCC and for all codecs), especially the PSNR based metrics which are consistently better than MSE or Hausdorff based metrics. Between the P2Point and Po2Plane PSNR there is no statistical difference, which means that both metrics achieve similar performance. Moreover, Pl2Plane metrics have the best performance for V-PCC decoded data (for *RColor* rendering). In summary, the statistical significance results allow confirming that the conclusions drawn before (Section V.B.1-3) are valid.

In summary, when all codecs and all renderings are considered, the Po2Point and Po2Plane PSNR metric values have the highest correlation with subjective scores. These metrics correspond to the ones previously selected by MPEG for the PCC Call for Proposals and currently used in common test conditions [42]. To the best of the authors knowledge, this is the first time that these metrics choice has been validated with MOS obtained with a well-defined procedure. From the results presented in this work, there is still significant room for improvement, especially if the goal is to achieve the same level of performance that past objective metrics (e.g. SSIM based) have obtained for 2D image and video representations.

## VI. FINAL REMARKS

The main objectives of this paper are to study the impact of the rendering process on the perceived quality of decoded PCs and the performance of available PC geometry objective metrics. To achieve these objectives, three representative PC coding solutions and three PC rendering solutions were used as well as a wide set of objective metrics. The subjective experiments suggest that geometric coding distortions can be masked by using the color attributes and (to a less extent) by surface reconstruction methods. Moreover, PC codecs produce distinct coding artifacts that have different impacts in terms of the final perceived quality, e.g. for PCL decoded data, geometric distortions are clearly visible for all rendering methods. Regarding the objective metrics evaluation, the results show that a careful selection of the objective metrics is necessary to have a reliable measure of the decoded PCs quality. Also, for some codecs and rendering solutions, the current metrics are not very reliable, e.g. for V-PCC coded data; this is rather critical since V-PCC is expected to become the first coding standard to be deployed in the market. Moreover, some of the objective metrics have a rather limited scope with significantly degraded accuracy, for some specific rendering solutions.

Regarding future work, a natural extension of this work is rendering with color attributes coded at different rates/qualities as it is critical to identify the best trade-off between geometry and color parts while maximizing the user perceived quality.

## VII. PC QUALITY ASSESSMENT: CHALLENGES AND WAYS FORWARD

The experimental results presented in this work allowed to derive several relevant conclusions to the PC coding and quality assessment fields. Establishing a bridge to those conclusions, this section identifies some challenges and suggests possible ways forward towards promoting advances on those fields, as follows:

**Advancing PC coding:** The study reported in this work shows that the number of decoded points plays a rather important role in perceived quality. When the density of points is increased during PC decoding, the perceived quality is also increased and, thus, better subjective scores can be obtained (as shown in Section V.B.1). This effect is clear for the MPEG V-PCC coding solution, which produces decoded PCs with artifacts less visible (and thus less annoying) as shown in sections III.C and IV.D (Fig. 8). However, many practical applications may not afford this increase on the size of the decoded PCs (due to memory and computational speed constrains), and thus new coding methods that tightly couple the decoding and the rendering processes are needed, e.g. PC rendering could render the polygons of the G-PCC *Trisoup* geometry representation directly. Moreover, PC

> REPLACE THIS LINE WITH YOUR PAPER IDENTIFICATION NUMBER (DOUBLE-CLICK HERE TO EDIT) <    15representations from which a varying number of decoded points can be sampled (e.g. from a set of triangles or 2D projection maps) could lead to more efficient ways of achieving gains in perceived quality, naturally without compromising the PC coding engine efficiency.

**Advancing PC subjective quality assessment:** The study reported in this work shows the importance of factoring the rendering process in the evaluation methodology. For example, if non-colored PCs are evaluated and a mesh-based rendering methodology is followed, some geometric artifacts (as shown in Section IV.D.2) may be masked, which should be avoided when the final rendering method is unknown. This is an important insight for future subjective studies, which assumes particular relevance since both JPEG and MPEG groups have been performing often this type of subjective evaluations and PC subjective quality assessment standards are not yet defined. Moreover, it is now clear from this work that the impact of color attributes in the overall perceived quality is high and masks geometry deformations (Section IV.D.1), which is not adequate for several applications, such as geographical information systems or automotive applications. Clearly, for these applications, fidelity is an important factor and thus subjective studies should include both geometry and geometry plus color subjective assessments to measure the different aspects of PC quality.

**Advancing objective PC metrics:** The study reported in this work shows that the final perceived quality not only depends on PC errors introduced by some processing step (in this case, coding) but also on the rendering process (see Section V.B.2 and Table VII). This way, metrics that explicit consider the way that rendering is performed are likely to perform better, for example, the distance between points (density) after projection could be considered for point-based rendering solutions. For mesh-based rendering solutions, suitable characterization of the surfaces using some statistical information, e.g. perceptually relevant surface-based and color-based features, can be extracted and used to predict the perceived visual quality. This may lead to much needed improvements, considering that objective metrics performance is rather poor for this type of rendering (as shown in Section V.B.3).

## REFERENCES

[1] H. Jung, H. Lee and C. E. Rhee, "Flexibly connectable light field system for free view exploration," *IEEE Trans. on Multimedia*, doi: 10.1109/TMM.2019.2934819, Aug. 2019.

[2] P. Rente, C. Brites, J. Ascenso and F. Pereira, "Graph-based static 3D point clouds geometry coding," *IEEE Trans. on Multimedia*, vol. 21, no. 2, pp. 284-299, Feb. 2019.

[3] J. Chen, C. Lin, P. Hsu and C. Chen, "Point cloud encoding for 3D building model retrieval," *IEEE Trans. on Multimedia*, vol. 16, no. 2, pp. 337-345, Feb. 2014.

[4] S. Park and S. Lee, "Multiscale representation and compression of 3-D point data," *IEEE Trans. on Multimedia*, vol. 11, no. 1, pp. 177-183, Jan. 2009.

[5] J. Sim and S. Lee, "Compression of 3-D point visual data using vector quantization and rate-distortion optimization," *IEEE Trans. on Multimedia*, vol. 10, no. 3, pp. 305-315, Apr. 2008.

[6] R. Mekuria, C. Tulvan and Z. Li. "Requirements for point cloud compression," *ISO/IEC MPEG N16330*, Geneva, Switzerland, Feb. 2016.

[7] T. Ebrahimi, S. Foessel, F. Pereira and P. Schelkens, "JPEG Pleno: toward an efficient representation of visual reality," *IEEE Multimedia*, vol. 23, no. 4, pp. 14-20, Oct.-Dec. 2016.

[8] S. Schwarz *et al.*, "Emerging MPEG standards for point cloud compression," *IEEE Journal on Emerging and Selected Topics in Circuits and Systems*, vol. 9, no. 1, pp. 133-148, Mar. 2019.

[9] MPEG 3DG and Requirements, "Call for proposals for point cloud compression," *ISO/IEC MPEG N16763*, Hobart, Australia, Apr. 2017.

[10] 3DG Group, "Text of ISO/IEC CD 23090-9 Geometry-based Point Cloud Compression," *ISO/IEC MPEG N18478*, Geneve, Switzerland, Mar. 2019.

[11] 3DG Group, "Text of ISO/IEC CD 23090-5: Video-based Point Cloud Compression," *ISO/IEC MPEG N18030*, Macau, China, Oct. 2018.

[12] M. Schütz and M. Wimmer. "High-quality point-based rendering using fast single-pass interpolation," *IEEE Digital Heritage*, Granada, Spain, Sept. 2015.

[13] P. Rosenthal and L. Linsen. "Image-space point cloud rendering," *Int. Conf. on Computer Graphics*, Istanbul, Turkey, Jun. 2008.

[14] E. Alexiou, E. Upenik and T. Ebrahimi, "Towards subjective quality assessment of point cloud imaging in augmented reality," *IEEE Workshop on Multimedia Signal Processing*, Luton, UK, Oct. 2017.

[15] E. Alexiou and T. Ebrahimi, "On the performance of metrics to predict quality in point cloud representations," *Applications of Digital Image Processing XL (SPIE 10396)*, San Diego, CA, USA, Sept. 2017.

[16] E. Alexiou and T. Ebrahimi, "On subjective and objective quality evaluation of point cloud geometry," *Int. Conf. on Quality of Multimedia Experience*, Erfurt, Germany, May 2017.

[17] E. Alexiou *et al.*, "Point cloud subjective evaluation methodology based on 2D rendering," *Int. Conf. on Quality of Multimedia Experience*, Sardinia, Italy, May 2018.

[18] E. Alexiou and T. Ebrahimi, "Impact of visualization strategy for subjective quality assessment of point clouds," *Int. Conf. on Multimedia & Expo Workshops*, San Diego, CA, USA, Jul. 2018.

[19] E. Alexiou *et al.*, "Point cloud subjective evaluation methodology based on reconstructed surfaces", *Applications of Digital Image Processing XL (SPIE 10752)*, San Diego, CA, USA, Sept. 2018.

[20] E. Alexiou and T. Ebrahimi, "Benchmarking of objective quality metrics for colorless point clouds," *Picture Coding Symposium*, San Francisco, CA, USA, Jun. 2018.

[21] J. Zhang, W. Huang, X. Zhu and J. Hwang, "A subjective quality evaluation for 3D point cloud models," *Int. Conf. on Audio, Language and Image Processing*, Shanghai, China, Jul. 2014.

[22] R. Mekuria, K. Blom and P. Cesar, "Design, implementation, and evaluation of a point cloud codec for tele-immersive video," *IEEE Trans. on Circuits and Systems for Video Technology*, vol. 27, no. 4, pp. 828-842, Apr. 2017.

[23] A. Javaheri, C. Brites, F. Pereira and J. Ascenso, "Subjective and objective quality evaluation of 3D point cloud denoising algorithms," *Int. Conf. on Multimedia & Expo Workshops*, Hong Kong, Jul. 2017.

[24] A. Javaheri, C. Brites, F. Pereira and J. Ascenso, "Subjective and objective quality evaluation of compressed point clouds," *IEEE Workshop on Multimedia Signal Processing*, Luton, UK, Oct. 2017.

[25] K. Christaki *et al.*, "Subjective visual quality assessment of immersive 3D media compressed by open-source static 3D mesh codecs," *Int. Conf. on Multimedia Modeling*, Thessaloniki, Greece, Jan. 2019.

[26] E. Dumic, C. R. Duarte and L. A. da Silva Cruz, "Subjective evaluation and objective measures for point clouds — State-of-the-art," *Int. Colloquium on Smart Grid Metrology*, Split, Croatia, May 2018.

[27] M. Pharr, W. Jakob and G. Humphreys, "Physically based rendering: from theory to implementation," *Morgan Kaufmann*, 2016.

[28] T. Akenine-Moller, E. Haines and N. Hoffman. "Real-time rendering," *AK Peters/CRC Press*, 3rd edition, ISBN 987-1-56881-424-7, 2018.

[29] V. S. Ramachandran, "Perception of shape from shading," *Nature*, vol. 331, no. 6152, pp. 163-166, Jan. 1988.

[30] R. L. Graham and P. Hell, "On the history of the minimum spanning tree Problem," *Annals of the History of Computing*, Jan.-Mar., 1985.

[31] M. Kazhdan, M. Bolitho and H. Hoppe, "Poisson surface reconstruction," *Eurographics Symp. on Geometry Processing*, Sardinia, Italy, Jun. 2006.

[32] A. Khatamian and H. R. Arabnia, "Survey on 3D surface reconstruction," *Journal of Information Processing Systems*, vol. 12, no. 3, pp. 338-357, Sep. 2016

[33] R. Zhang, P.-S. Tsai, J. E. Cryer and M. Shah, "Shape-from-shading: a survey," *IEEE Trans. on Pattern Analysis and Machine Intelligence*, vol. 21, no. 8, pp. 690-706, Aug. 1999.

[34] D. F. Fouhey, A. Gupta and A. Zisserman, "From images to 3D shape attributes," *IEEE Trans. on Pattern Analysis and Machine Intelligence*, vol. 41, no. 1, pp. 93-106, Dec. 2017.

[35] D. Girardeau-Montaut, "Cloud Compare—3D point cloud and mesh processing software," Available: http://www.cloudcompare.org/.

[36] R.B. Radu and S. Cousins, "3D is here: Point Cloud Library (PCL)," *IEEE Int. Conf. on Robotics and Automation*, Shanghai, China, May 2011.

[37] J. Kammerl, N. Blodow, R. B. Rusu, S. Gedikli, M. Beetz and E. Steinbach, "Real-time compression of point cloud streams," *Int. Conf. on Robotics and Automation*, Saint Paul, MN, USA, May 2012.

[38] G. Nigel and N. Martin, "Range encoding: an algorithm for removing redundancy from a digitized message," *Video and Data Recording Conf.*, Southampton, UK, Jul. 1979.